# Franck-Condon blockade in suspended carbon nanotube quantum dots

Renaud Leturcq[1,2,*], Christoph Stampfer[1,3,*], Kevin Inderbitzin[1], Lukas Durrer[3], Christofer Hierold[3], Eros Mariani[4], Maximilian G. Schultz[4], Felix von Oppen[4] & Klaus Ensslin[1]

[1] *Laboratory for Solid State Physics, ETH Zurich, 8093 Zurich, Switzerland*

[2] *Institut d'Electronique de Microélectronique et de Nanotechnologie, CNRS-UMR 8520, Department ISEN, Avenue Poincaré, BP 60069, 59652 Villeneuve d'Ascq Cedex, France*

[3] *Micro and Nanosystems, Department of Mechanical and Process Engineering, ETH Zurich, 8092 Zurich, Switzerland*

[4] *Institut für Theoretische Physik, Freie Universität Berlin, Arnimallee 14, 14195 Berlin, Germany*

[*] *These authors contributed equally to this work*

**Understanding the influence of vibrational motion of the atoms on electronic transitions in molecules constitutes a cornerstone of quantum physics, as epitomized by the Franck-Condon principle[1,2] of spectroscopy. Recent advances in building molecular-electronics devices[3] and nanoelectromechanical systems[4] open a new arena for studying the interaction between mechanical and electronic degrees of freedom in transport at the single-molecule level. The tunneling of electrons through molecules or suspended quantum dots[5,6] has been shown to excite vibrational modes, or vibrons[7-9,6]. Beyond this effect, theory predicts that strong electron-vibron coupling dramatically suppresses the current flow at low biases, a collective behaviour known as Franck-Condon blockade[10]. Here we show measurements on quantum dots formed in suspended single-wall carbon nanotubes revealing a remarkably large electron-vibron coupling and, due to the high quality and unprecedented tunability of our samples, admit a quantitative analysis of vibron-mediated electronic transport in the regime of strong electron-vibron coupling. This allows us to unambiguously demonstrate the Franck-Condon blockade in a suspended nanostructure. The large observed electron-vibron coupling could ultimately be a key ingredient for the detection of quantized mechanical motion[11,12]. It also emphasizes the unique potential for nanoelectromechanical device applications based on suspended graphene sheets and carbon nanotubes.**

In a polar semiconductor, a conduction electron deforms the surrounding lattice to form a polaron state[13]. The formation of this quasi-particle, by combining an electron and a cloud of lattice vibrations, or phonons, strongly influences the transport properties. The possibility for localization of strongly coupled polarons was suggested by Landau more than 70 years ago[13]. Recently, Koch *et al.* predicted that a related trapping of heavy polarons can occur in a quantum dot (QD) formed in a mechanically suspended nanostructure[10]. In such a nanoelectromechanical system (NEMS), the vibrational modes of the nanostructure can be strongly affected by the presence of electrons in the QD, as they deform the embedding medium. For strong electron-phonon coupling, the deformation effectively blocks electronic transport, termed Franck-Condon (FC) blockade. By analysing electronic transport through a suspended carbon nanotube (CNT) quantum dot over a wide range of electronic states, we are able to highlight *generic* features of vibron-assisted electronic transport, and unambiguously confirm the FC blockade scenario.

Scanning electron microscope images and a scheme of our suspended CNT quantum dot device are shown in Figs. **1a**, **1b** and **1c**. The CNT is electrically and mechanically connected to both source (S) and drain (D) contacts, while the central electrode acts as a suspended top-gate (TG). A quantum dot in the CNT is formed between defects[14], which are presumably created during the release process and act as local barriers. The double top- and back-gate configuration allows us to determine the location of the QD below the TG (see Fig. **1c** and supplementary material).

The exceptional quality of our sample, which is crucial for this investigation, is revealed by the well resolved observation of multiple four-fold degeneracies of electronic states in the large-scale presentation of our data (Fig. **1d**). This stems from the combined spin and valley degeneracies in clean CNT[15-17], and allows us to fully characterize the electronic properties of our device, including the electronic confinement energy $\Delta E_{elec} = 6.8 \pm 2.5$ meV (see supplementary material).

At a finer energy scale (see Fig. **2**), all the probed Coulomb diamonds show several striking generic features. 1) Quasi-periodic lines running parallel to the edges of the Coulomb diamonds, corresponding to excited states, appear ubiquitously. 2) For increasing temperatures, we observe the appearance of absorption sideband peaks within the Coulomb blockaded regions. 3) At the edge of the diamonds, there is a pronounced current suppression compared to the current magnitude for excited states. 4) Most diamonds show a significant apparent shift of their tip between positive and negative bias. 5) Finally, all the probed diamonds show ubiquitous negative differential conductance (NDC) appearing in between the excited states. We will show that most of these features are generic for transport-induced vibron excitations and are signatures of the strong electron-vibron coupling in our system.

Quasi-periodic excited states have been reported in previous experiments on suspended quantum dots. They have been interpreted as being due to the excitation of quantized phonons in the nanostructure, called vibrons[7-9,6]: when the bias voltage matches a multiple of the energy of a vibron, $\hbar\omega_0$, transport is enhanced by *emission* of a vibron. This interpretation is confirmed in our data since the energy spacing between these excited states $\Delta E_{vib}$ =0.8 ± 0.2 meV is much lower than the measured electronic level spacing. In addition, Fig. **2b** shows that this energy is constant over a wide range of gate voltages and does not depend on the electronic structure of the quantum dot.

The bosonic origin of the excitations is further demonstrated by the temperature dependence of the data. As temperature is increased, we observe additional excited states appearing within the Coulomb blockade regime (Figs. **2c** to **2f**), similar to anti-Stokes resonances in Raman spectroscopy. This behaviour, not reported in previous experiments, is expected for vibronic excited states since higher vibrational states become populated at elevated temperature. As a consequence, electronic transport is enabled even in the Coulomb-blockade regime by *absorption* of a vibron. The quantitative analysis of the peak magnitude as a function of temperature confirms the bosonic origin of the absorption peak (Fig. **2g**).

The very observation of vibrational sidebands is a signature of rather strong electron-vibron couplings in our CNT quantum dot. This is further highlighted by the current suppression at low bias detected in all the probed diamonds.

Indeed, our data unambiguously confirm that this is a direct demonstration of the so-called FC blockade[10]. When tunnelling onto the dot, the electron shifts the equilibrium coordinate of the vibron oscillator by an amount proportional to the electron-vibron coupling (Fig. **3a**). The transition probability of this tunnelling process is proportional to the square of the overlap between the vibronic wave functions before and after tunnelling. In spectroscopy, this principle is well known as FC principle[1,2] and explains the magnitude of peaks in absorption spectra. Since the overlap of low-lying vibronic states is exponentially sensitive to this geometrical displacement, the ground-state-to-ground-state transition for vibrons is exponentially suppressed for strong electron-vibron coupling. This leads to the FC blockade of the current at low bias. At the same time, the overlap between the vibronic ground state with highly excited states, whose wave functions are widely spread in space, is still significant. This allows for tunnelling while exciting vibrons, as long as the bias is large enough to compensate for the vibronic energy difference, causing the observed vibrational sidebands.

This interpretation provides a quantitative understanding of the experimental data. The apparent shift of the Coulomb diamond tips for positive and negative bias observed in many cases in our experiment (Fig. **3c**) and others[5], is explained by the asymmetric tunnel coupling of the QD with source and drain (Fig. **3e**). The smaller tunnel coupling determines the current limiting tunnelling process, and energetic considerarions (sketched in Fig. **3e**) imply that the edge of the diamond is due to ground-state-to-ground-state transitions on one side and due to ground-state-to-excited-state transitions for the other. As a result, the solution (see Fig. **3d**) of the corresponding rate equations in Fig. **3e** reproduces the experiment in Fig. **3c** quite accurately.

In order to extract the electron-vibron coupling $g$ despite the above asymmetry, we study the differential conductance at the gate voltage corresponding to the diamond tip (Figs. **3f** and **3g**). For equilibrated phonons, the Franck-Condon theory predicts differential conductance peaks following the progression:

$$\left(\frac{dI}{dV}\right)_n^{max} \propto |M_{0\to n}|^2 \propto \frac{e^{-g}g^n}{n!} \tag{1},$$

with $n$ the difference in vibron quantum numbers and $|M_{0n}|$ the overlap between vibronic states with 0 and $n$ vibrons. By fitting the maxima of dI/dV with this expression (Figs. **3f** and **3g**), we find very good agreement for most measured diamonds. Our measurement not only shows a current suppression at low bias[5,6] but rather matches the whole progression for excited states, proving global consistency with the Franck-Condon picture. We deduce a large value for the electron-vibron coupling parameter, $g$ = 3.3 ±

0.9 over the whole set of Coulomb diamonds. This is about a factor 3 larger than in previous measurements[6].

In order to obtain a theoretical understanding of the large electron-vibron coupling constant $g$ in the QD, we consider the intrinsic electron-phonon coupling of the underlying CNT. This turns out to be dominant compared to previously considered extrinsic electrostatic coupling mechanisms[6,18]. The coupling originates from (a) the modified hopping associated with changes in the C-C bond lengths and (b) the deformation potential due to local area variations; the deformation potential coupling constant $g_D \approx$ 30 eV is about an order of magnitude larger than the hopping-induced one[19,20].

When combined with symmetry considerations, the existence of these two coupling mechanisms has several important consequences for the coupling strengths of the various CNT vibron modes. Longitudinal stretching modes (LSM) and radial breathing modes couple most strongly. Their associated lattice deformations involve local area variations and produce a linear shift of the electronic energies, resulting in a linear deformation potential interaction. Twist modes also couple linearly, but they are area preserving and couple only via the weaker hopping-induced mechanism. Finally, bending modes have only a weak quadratic interaction since, for symmetry reasons, the electronic energy does not depend on the sign of the associated deformation of the CNT.

We can thus identify the relevant vibron mode probed in our data. Indeed, of the strongly coupled vibrons, only the LSM is in a frequency range compatible with our experimental observations (see Fig. **2b**), as discussed previously[6]. The relevant dimensionless coupling constant $g$ for the occurrence of the Franck-Condon blockade is given by (the square of) the shift of the equilibrium coordinate measured in units of the amplitude (oscillator length) of the vibronic quantum fluctuations. We compute this quantity in the framework of the effective Dirac Hamiltonian for the low-energy electronic properties of CNTs, combined with the theory of elasticity for the description of the vibron mode (see details in the supplementary material). Using literature values for the elastic constants of graphene[20], this calculation yields a maximal coupling of

$$g_{max}^{(LSM)} \simeq \frac{10}{L_\perp [nm]} \qquad (2)$$

for the lowest LSM. An important consequence of our analysis is that the coupling constant is strongly sample dependent. It is inversely proportional to the CNT circumference $L_\perp$, and proportional to the electronic matrix element of the deformation potential. The matrix element vanishes for the lowest longitudinal stretching mode when electrons and vibrons are confined to the same region of the CNT. It becomes maximal when the electronic wave function is sharply localized around a region of maximum strain, yielding the estimate in Eq. (2).

AFM measurements indicate that our CNT typically have a circumference of a few nanometers. Thus, our value for $g$ given in Eq. (2) is consistent with the observation of the Franck-Condon blockade which requires $g > 1$. At the same time, the coupling constants extracted from experiment are slightly larger than our estimate. Other observations also indicate that experiments exhibit more pronounced vibronic effects in transport than theory suggests: vibrational sidebands are accompanied by negative differential conductance, as already reported[6], while theory predicts a step-like $I$-$V$ characteristic. In addition, the experiment shows vibrational absorption sidebands within the Coulomb blockade diamonds which are absent in theoretical simulations. We have checked that these intriguing enhancements are not related to asymmetric coupling of the dot to the leads (see supplementary material). The underlying reasons are not understood at present.

**Methods**

Devices were fabricated from highly doped silicon wafer substrates covered by 200 nm silicon oxide. Single-walled carbon nanotubes (SWNTs) were grown randomly on the oxide substrate by CVD based on dispersed Ferritin catalysts[21]. Electron beam lithography was used to pattern metallic electrodes and gate structures (2/30 nm Cr/Au) around (selected) individual SWNTs (Fig. **1b**). Finally diluted HF (4% for 5 min) etching followed by critical point drying completes the device fabrication (Fig. **1c**). It is crucial that the Cr layer oxidises after the etching and drying including exposing to environmental air in order to form the top-gate oxide. For a more detailed process description we refer to Refs. 22 and 23. Measurements were performed in a variable temperature $^4$He cryostat at a base temperature of $T$ = 1.3 K. We have measured the two-terminal conductance through the graphene SET device by applying a symmetric DC bias voltage $V_{SD}$ while measuring the current through the SET device with a resolution better than 10 fA. For differential conductance measurements a small AC bias, $V_{ac}$ = 48 µV was superimposed on $V_{SD}$ and the differential conductance was measured with lock-in techniques. Measurements have been done in two different devices, with quantum dots formed at different locations along the CNT. All configurations show similar results as the one presented here.

1. Franck, J. Elementary processes of photochemical reactions. *Transactions of the Faraday Society* **21**, 536-542 (1926).

2. E. Condon. A theory of intensity distribution in band systems. *Phys. Rev.* **28**, 1182-1201 (1926).

3. Joachim, C., Gimzewski, J. K. & Aviram, A. Electronics using hybrid-molecular and mono-molecular devices. *Nature (London)* **408**, 541-548 (2000).

4. Ekinci, K. L. & Roukes, M. L. Nanoelectromechanical systems. *Rev. Sci. Instr.* **76**, 061101 (2005).

5. Weig, E. M. *et al.* Single-Electron-Phonon Interaction in a Suspended Quantum Dot Phonon Cavity. *Phys. Rev. Lett.* **92**, 046804 (2004).

6. Sapmaz, S., Jarillo-Herrero, P., Blanter, Y. M., Dekker, C. & van der Zant, H. S. J. tunneling in Suspended Carbon Nanotubes Assisted by Longitudinal Phonons. *Phys. Rev. Lett.* **96**, 026801 (2006).

7. Park, H. *et al.* Nanomechanical oscillations in a single-C60 transistor. *Nature (London)* **407**, 57-60 (2000).

8. Yu, L. H. *et al.* Inelastic Electron Tunneling via Molecular Vibrations in Single-Molecule Transistors. *Phys. Rev. Lett.* **93**, 266802 (2004).

9. Pasupathy, A. N. *et al.* Vibration-Assisted Electron Tunneling in C140 Transistors. *Nano Lett.* **5**, 203-207 (2005).

10. Koch, J. & von Oppen, F. Franck-Condon Blockade and Giant Fano Factors in Transport through Single Molecules. *Phys. Rev. Lett.* **94**, 206804 (2005).

11. Knobel, R. G. & Cleland, A. N. Nanometre-scale displacement sensing using a single electron transistor. *Nature (London)* **424**, 291-293 (2003).

12. LaHaye, M. D., Buu, O., Camarota, B. & Schwab, K. C. Approaching the Quantum Limit of a Nanomechanical Resonator. *Science* **304**, 74-77 (2004).

13. Landau, L. D. Electron motion in crystal lattice. *Phys. Z. Sowjet.* **3**, 664-665 (1933).

14. Postma, H. W. C., Teepen, T., Yao, Z., Grifoni, M. & Dekker, C., Carbon Nanotube Single-Electron Transistors at Room Temperature. *Science* **293**, 76-79 (2001).

15. Liang, W., Bockrath, M. & Park, H. Shell Filling and Exchange Coupling in Metallic Single-Walled Carbon Nanotubes. *Phys. Rev. Lett.* **88**, 126801 (2002).

16. Buitelaar, M. R., Bachtold, A., Nussbaumer, T., Iqbal, M. & Schönenberger, C. Multiwall Carbon Nanotubes as Quantum Dots. *Phys. Rev. Lett.* **88**, 156801 (2002).

17. Sapmaz, S. *et al.* Electronic excitation spectrum of metallic carbon nanotubes. *Phys. Rev. B* **71**, 153402 (2005).

18. Braig, S. & Flensberg, K. Vibrational sidebands and dissipative tunneling in molecular transistors. *Phys. Rev. B* **68**, 205324 (2003).

19. Woods, L. M. & Mahan, G. D., Electron-phonon effects in graphene and armchair (10,10) single-wall carbon nanotube. *Phys. Rev. B* **61**, 10651-10663 (2000).

20. Suzuura, H. & Ando, T. Phonons and electron-phonon scattering in carbon nanotubes. *Phys. Rev. B* **65**, 235412 (2002).

21. Durrer, L., Hebling, T., Zenger, C., Junger, A., Stampfer & C., Hierold, C. SWNT growth by CVD on Ferritin-based iron catalyst nanoparticles towards CNT sensors. *Actuators and Sensors and Actuators B: Chemical* **132**, 485-490 (2008).

22. Stampfer, C. *et al.* Nano-Electromechanical Displacement Sensing Based on Single-Walled Carbon Nanotubes. *Nano Lett.* **6**, 1449-1453 (2006).

23. Stampfer, C., Jungen, A. & Hierold, C. Fabrication of discrete nanoscaled force sensors based on single-walled carbon nanotubes. *IEEE Sensors Journal* **6**, 613-617 (2006).

24. Dresselhaus, M. S. & Eklund, P. C. Phonons in carbon nanotubes. *Adv. Phys.* **49**, 705-814 (2000).

25. Kouwenhoven, L. P. *et al.* Electron transport in quantum dots. In Sohn, L. L., Kouwenhoven, L. P. & Schön, G. (eds.) *Mesoscopic Electron Transport*, NATO ASI Ser. E 345, 105-214 (Kluwer, Dordrecht, 1997).

We thank K. Kobayashi, L. Wirtz, D. Obergfell, M. Muoth, M. L. Roukes, S. Ilani for helpful discussions. This work was supported by the TH-18/03-1 Grant and the Swiss National Science Foundation via NCCR Nanoscience, as well as Sfb658 and SPP 1243. Sample fabrication was performed at the ETH FIRST and CLA Laboratories.

Correspondence and requests for materials should be addressed to R.L. (email: renaud.leturcq@isen.iemn.univ-lille1.fr).

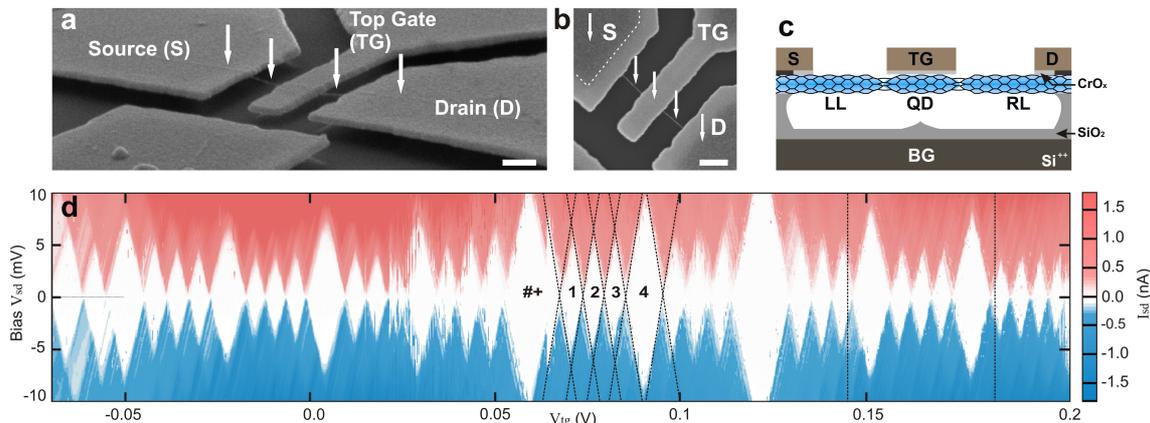

**Figure 1: Characterization of the suspended carbon nanotube quantum dot. a** Electron microscope micrograph in tilted view of the suspended carbon nanotube (white arrows) with the source (S) and drain (D) electrodes and the central top-gate (TG) electrode. **b** Top view of the device. The scale bar in **b** and **c** is 200 nm. **c** Scheme of the quantum dot device formed in the suspended carbon nanotube. As shown by the charge stability diagrams (see supplementary materials), a quantum dot is formed under the top-gate, separated from the other parts of the nanotube through tunnel barriers. Both the suspended top-gate and the underlying back-gate (BG) can be used to tune the electronic properties of the quantum dot and the leads. **d** Source-drain current through the quantum dot measured as a function of the bias voltage $V_{sd}$ and the top-gate voltage $V_{tg}$, adjusting the back-gate voltage simultaneously in order to keep the average chemical potential in the leads constant (see supplementary informations): $V_{bg} = -0.7576 \times V_{tg} + 1.6784$ V. Diamond-shape regions of suppressed current are characteristic of Coulomb blockade (so-called Coulomb diamonds), while the four-fold periodicity is characteristic of quantum dots formed in clean carbon nanotubes.

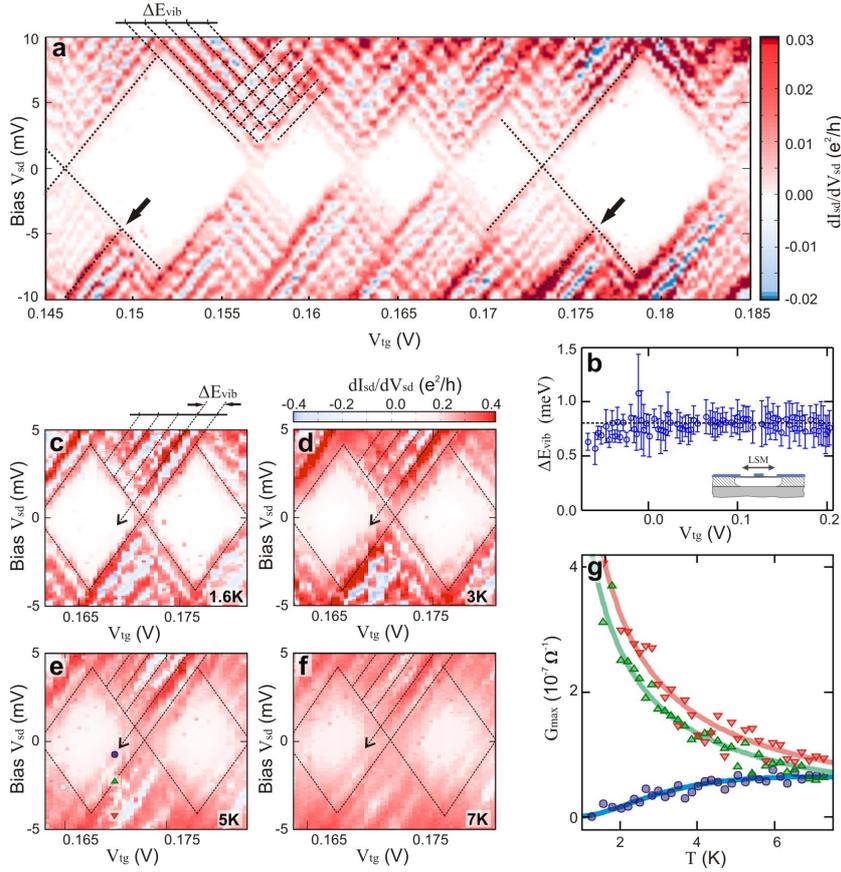

**Figure 2: Evidence and temperature dependence of vibron-assisted transport. a** Differential conductance d$I_{sd}$/d$V_{sd}$ for a subset of the Coulomb diamonds shown in Fig. **1d**, showing the quasi-periodic vibronic excited states (see dashed lines). The arrows point to electronic excited states, visible at higher energy. **b** Energy spacing of the vibronic excited states for the whole gate voltage range shown in Fig. **1d**. For each point, the energy spacing is averaged over half of a Coulomb diamond, corresponding to a given electronic state. The horizontal dashed line represents the average value over all points, $\Delta E_{vib}$ = 0.8 ± 0.2 meV. This energy is compatible with the longitudinal stretching mode, for which $\Delta E_{vib} = \hbar v_{ph} q$, with $q = n\pi/L$ for the $n^{th}$ vibron mode in the doubly clamped nanotube and $L$ the confinement length. Using the value of the stretching phonon group velocity in clean suspended CNT, $v_{st} \approx 2.4 \times 10^4$ m/s[24], we deduce the characteristic length for the first vibron mode $L \approx 65$ nm, of the same order of magnitude as the full suspended parts of the nanotube and in reasonable agreement with earlier experiments[6]. **c-f** Coulomb diamonds measured in the same region of gate voltage (same electronic state) for different temperatures, 1.6 K (**c**), 3 K (**d**), 5 K (**e**) and 7 K (**f**). As the temperature increases, additional conductance peaks appear in the Coulomb blockaded regions. In **e**, the green up triangle and red down triangle mark the positions of conductance peaks corresponding to tunneling through the ground state and the first vibronic excited state (emission peak), and the blue circle to the position of a conductance peak due to the absorption of a vibron (absorption peak). **g** Maximum conductance $G_{max}$ for tunneling through the ground state (green up triangle), the emission peak (red down triangle) and the absorption peak (blue circles), corresponding to points marked in panel **e**. For the tunneling through the ground state and the emission peak, we use as a fit $G_{max} \propto 1/k_B T$ (see green and red curves), as expected for the derivative of the Fermi distribution in the quantum Coulomb blockade regime[25]. For the absorption peak, we find a good fit with the product of the equilibrium population of the phonon states, given by the Bose distribution, $1/(\exp(\hbar\omega_0/k_B T)-1)$, and the derivative of the Fermi distribution, giving $G_{max} \propto 1/k_B T \times 1/(\exp(\hbar\omega_0/k_B T)-1)$ (see blue curve). From the fit to the experimental data we determine $\hbar\omega_0$ = 0.96 ± 0.08 meV, which is consistent with the energy spacing of the vibrational excited states ($\Delta E_{vib.}$ = 0.88 ± 0.15 meV at this value of the gate voltage).

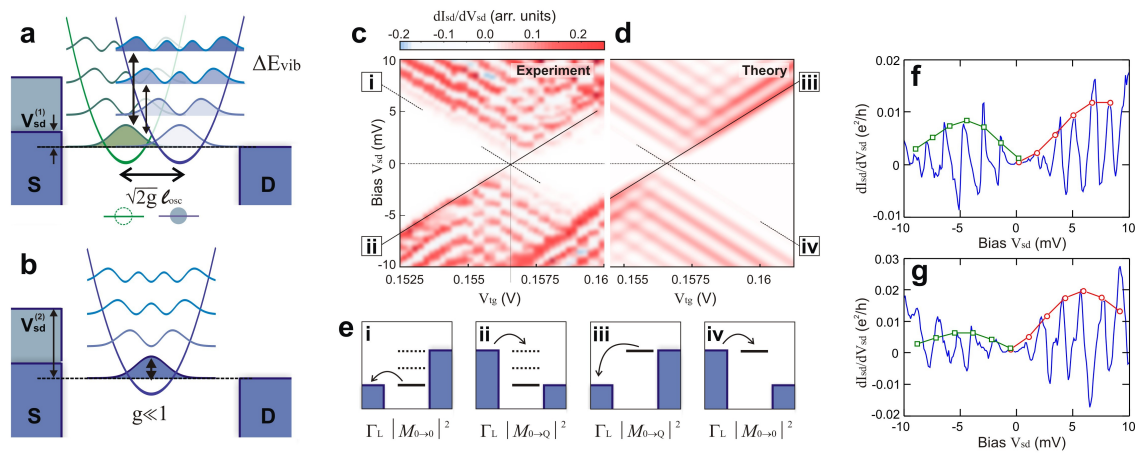

**Figure 3: Theory and analysis of Franck-Condon blockade. a,b** Schematic description of the displacement of the vibron potential for strong (**a**, $g \gg 1$) and weak (**b**, $g \ll 1$) electron-vibron coupling $g$, with no electron (green curve) and one electron (blue curve) in the dot. **a** The shift for $g \gg 1$ suppresses the transitions between low lying vibronic states and thus the current in the low bias regime ($V_{sd}^1$), causing the Franck-Condon blockade. In parallel, it opens transitions to excited states in the high bias regime ($V_{sd}^2$), yielding vibrational sidebands. In contrast, for $g \ll 1$ the shift is essentially absent, allowing ground state to ground state vibronic transitions. In parallel, the vibronic ground state in the absence of the electron is essentially orthogonal to all excited states in the presence of the electron. This suppresses the transition probabilities to highly excited states and the vibrational sidebands. **c** Zoom into a part of the Coulomb diamonds (around $V_{tg}$ = 0.1570 V) showing the suppression of the low-bias transport and the apparent shift of the Coulomb diamond tip at positive bias with respect to that at negative one, and simulated Coulomb diamonds in the same conditions as in the experiment (with $k_B T = 0.15 \times \hbar\omega_0$) for strongly asymmetric tunnel coupling between the leads and the dot, $\Gamma_R \gg \Gamma_L$. The dominant rate limiting processes are sketched for the extremal lines (dashed in the figure) of the Coulomb diamond. For the extremal line with negative slope the relevant process implies tunneling through the left barrier with no vibronic excitation (i and iv). These processes lead to a current proportional to $\Gamma_L|M_{0\to 0}|^2$ involving the Franck-Condon overlap between vibronic ground states $M_{0\to 0}$. Thus the Franck-Condon suppression of vibronic transitions manifests itself also at large bias for strongly asymmetric tunneling barriers. The extremal line with positive slope involves tunneling through the left lead as well, but as $V_{sd}$ increases inelastic excitation of vibrons can be energetically allowed (ii and iii). These processes are associated with vibronic transitions involving $M_{0\to Q}$, where $Q = \text{Int}[eV_{sd}/\hbar\omega_0]$ is the maximum allowed vibronic quantum number (as long as $M_{0\to Q}$ increases for increasing $Q$). The Franck-Condon mechanism does not suppress these matrix element for large $Q$, resulting in an increasing differential conductance peak at large bias, proportional to $\Gamma_L|M_{0\to Q}|^2$. **d** Differential conductance measured for $V_{tg}$ = 0.0678 V. Representative fits of the maxima with the Franck-Condon progression (see text) allows us to extract the parameters $g$ = 3.5 (negative bias voltage, green squares) and $g$=5.0 (positive bias voltage, red circles). Peaks that deviate strongly from the progression at large bias voltage are due to electronic excited states. **e** Same as in panel **d**, for $V_{tg}$ = 0.1570 V, giving $g$ = 3.0 (negative bias voltage, green squares) and $g$ = 4.5 (positive bias voltage, red circles).